# Two-Stage Active Distribution Network Voltage Control via LLM-RL Collaboration: A Hybrid Knowledge-Data-Driven Approach

Xu Yang, *Graduate Student Member, IEEE*, Chenhui Lin, *Senior Member, IEEE*
Xiang Ma, Dong Liu, Ran Zheng, Haotian Liu, *Member, IEEE*, and Wenchuan Wu, *Fellow, IEEE*

*Abstract*—The growing integration of distributed photovoltaics (PVs) into active distribution networks (ADNs) has exacerbated operational challenges, making it imperative to coordinate diverse equipment to mitigate voltage violations and enhance power quality. Although existing data-driven approaches have demonstrated effectiveness in the voltage control problem, they often require extensive trial-and-error exploration and struggle to incorporate heterogeneous information, such as day-ahead forecasts and semantic-based grid codes. Considering the operational scenarios and requirements in real-world ADNs, in this paper, we propose a hybrid knowledge-data-driven approach that leverages dynamic collaboration between a large language model (LLM) agent and a reinforcement learning (RL) agent to achieve two-stage voltage control. In the day-ahead stage, the LLM agent receives coarse region-level forecasts and generates scheduling strategies for on-load tap changer (OLTC) and shunt capacitors (SCs) to regulate the overall voltage profile. Then in the intra-day stage, based on accurate node-level measurements, the RL agent refines terminal voltages by deriving reactive power generation strategies for PV inverters. On top of the LLM-RL collaboration framework, we further propose a self-evolution mechanism for the LLM agent and a pretrain-finetune pipeline for the RL agent, effectively enhancing and coordinating the policies for both agents. The proposed approach not only aligns more closely with practical operational characteristics but also effectively utilizes the inherent knowledge and reasoning capabilities of the LLM agent, significantly improving training efficiency and voltage control performance. Comprehensive comparisons and ablation studies demonstrate the effectiveness of the proposed method.

## I. Introduction

RESPONDING to climate emergencies, pollution from fossil fuels, and carbon neutrality commitments, renewable energy resources, particularly distributed photovoltaics (PVs), are being increasingly integrated into distribution networks, transforming them into active distribution networks (ADNs) [1], [2]. While this integration substantially reduces carbon emissions at the distribution level, the growing penetration of PV has introduced significant operational challenges, notably voltage violations at the network extremities. Also, the inherent intermittency and stochastic nature of PV generation further exacerbate voltage fluctuations, power imbalances, and overall power quality degradation, posing serious threats to ADN stability and security [3], [4].

In this context, traditional passive voltage control methods are no longer sufficient to meet the demands of ADN operation. The ADN operators must adopt proactive voltage control approaches that jointly utilize mechanical assets such as on-load tap changer (OLTC) and shunt capacitors (SCs), as well as flexible resources like PV inverters capable of dynamic reactive support. Due to temporal coupling constraints and the requirements for manual on-site intervention, OLTC and SCs usually necessitate day-ahead scheduling based on the forecasted information or historical records. In contrast, during intra-day operation, the fast reactive power regulation capability of PV inverters can be leveraged to enable rapid control based on real-time measurements.

A variety of data-driven approaches, primarily based on deep reinforcement learning (RL), have been developed to enable efficient resolution of the aforementioned voltage control problem [5]-[14]. These methods fundamentally operate by allowing an intelligent RL agent to iteratively interact with the external environment and enhance its policy based on the received feedback. For example, researchers in [5]-[7] proposed RL-based voltage control methods for ADNs, optimizing voltage profiles by utilizing internal distributed energy resources. Researchers in [8]-[11] extended this paradigm into multi-agent settings, where different system regions or resource clusters are assigned to individual RL agents that collaborate to achieve overall voltage optimization. Additionally, researchers in [12]-[14] addressed the coordination of different equipment operating on two timescales, enabling hierarchical voltage regulation across various temporal horizons.

Although the above RL-based data-driven methods have demonstrated effective and promising performance, their practical application to real-world ADN voltage control, especially in the day-ahead stage, still faces four key

This work was supported in part by the State Grid Zhejiang Electric Power Company Limited Science and Technology Project "Research on Key Technologies for Cooperative Regulation of Flexible Resources in Active Distribution Networks Based on Highly Safe and Scalable Artificial Intelligence" under Grant 5211JH250006 *(Corresponding author: Wenchuan Wu).*

Xu Yang, Chenhui Lin, Haotian Liu, and Wenchuan Wu are with the Sichuan Energy Internet Research Institute, Tsinghua University, Chengdu 610213, China.

Xiang Ma, Dong Liu, and Ran Zheng are with the State Grid Zhejiang Electric Power Co., Ltd., Jinhua Power Supply Company, Jinhua 321017, China.

challenges:

1) *Incomplete and heterogeneous information.* The effectiveness of data-driven methods typically hinges on the availability of complete and high-quality data, which corresponds to accurate node-level measurements in the context of voltage control problem. However, limited by the resolution of current meteorological forecasts and the accuracy of PV/load predictions, only coarse-grained information, i.e., hourly region-level forecasts are available in the day-ahead stage. Moreover, some unstructured and heterogeneous information such as historical records and voltage reports, may also be provided in the day-ahead stage. Processing and integrating such incomplete and heterogeneous data pose significant challenges for existing RL algorithms.

2) *Semantics-based operational constraints.* The scheduling of ADN mechanical assets usually involves numerous semantic-based grid codes, such as: "The number of daily adjustments of an OLTC is subject to an upper limit," or "SCs can only be committed during a specific time window each day due to the requirement for manual switching." These grid codes are essentially operational constraints on the agent's action space. However, incorporating such constraints into an RL policy typically requires elaborate techniques, such as penalty mechanism [14] or action masking [15], which reduces the training efficiency of the RL agent.

3) *Instructions adapted to long-tail requirements.* Since mechanical assets may encounter unforeseen events such as faults or maintenance, ADN operators may occasionally issue operational instructions described in natural language. Therefore, the agent's policy must be capable of adapting to such instructions. However, RL agents are inherently unable to process emergent natural-language instructions and exhibit poor generalization to unseen scenarios, which can result in performance degradation or even infeasible control actions.

4) *Ambiguous policy improvement direction.* Finally, the RL agent relies solely on a scalar reward signal to guide its training process. However, such a scalar reward carries very limited information and fails to provide a clear direction for policy evolution. For example, when the overall voltage is low, the optimal action, such as raising the OLTC tap position, is intuitively straightforward. Yet, the RL agent must undergo extensive exploration to discover this simple rule, resulting in lower training efficiency or worse performance.

The rapid advancement of large language models (LLMs) in recent years has initiated widespread applications across various domains [16]-[18] and offered innovative solutions to the above challenges. On the one hand, LLMs possess inherent natural-language processing and strong information integration abilities, enabling them to receive, analyze, and process incomplete and heterogeneous information. On the other hand, LLMs embed a vast amount of prior knowledge and exhibit powerful reasoning capabilities, allowing them to comprehend semantic-based operational constraints and instructions and generate valid responses accordingly. Building on these strengths, some pioneering studies have begun applying LLMs to the power system dispatch domain [19]-[21], while others have investigated how LLMs can assist the training process of RL agents [22]-[24]. In these studies, LLMs are generally limited to auxiliary subtasks such as code programming or function synthesis and are primarily treated as a natural-language interface, rather than being used for policy formulation. Furthermore, research on how to improve LLM policies remains comparatively scarce.

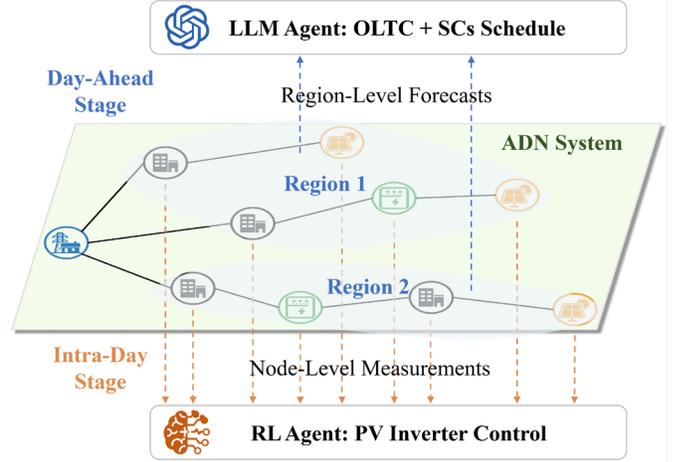

**Fig. 1.** Overall scheme of the proposed approach.

Considering practical requirements and operational scenarios in ADN voltage control, we propose a hybrid knowledge-data-driven approach that combines respective advantages of LLM and RL. As shown in Fig. 1, in the day-ahead stage, the ADN operator receives hourly region-level forecasts and utilizes an LLM agent to generate scheduling strategies for the OLTC and SCs. Then in the intra-day stage, the ADN operator receives accurate node-level measurements and utilizes an RL agent to generate real-time reactive power generation strategies for PV inverters. On the one hand, this setup aligns well with the two-stage nature of voltage control problem: in the day-ahead stage, coarse information is used to dispatch mechanical assets for regulating the overall voltage profile, while in the intra-day stage, fine-grained information enables PV inverters to refine terminal voltages. On the other hand, the proposed approach synergistically leverages the complementary advantages of LLM and RL, which not only exploit the LLM's information processing capabilities and embedded knowledge to enhance policy learning efficiency but also RL's precise computation and rapid response to optimize control performance.

In essence, different from the existing LLM-assisted RL paradigm, the proposed approach addresses a complex multi-stage constrained optimization problem through LLM-RL collaboration using different aspects of information. From the LLM's perspective, the RL agent serves as a callable specialized tool capable of performing inverter reactive control tasks, thereby extending the LLM's capability boundaries. Conversely, from the RL's perspective, the LLM agent constitutes an integral part of the external environment that processes heterogeneous or semantic information, effectively reducing problem complexity and shrinking the action space that needs to be explored, thus accelerating RL policy convergence.

Building upon the proposed LLM-RL collaboration framework, the key challenge lies in how to improve and coordinate the policies of the LLM agent and the RL agent. To this end, we propose a Reflexion [25]-based self-evolution mechanism for the LLM policy, which continuously enhances its decision-making through environmental feedback and knowledge base updates. Meanwhile, we also propose a pretrain-finetune pipeline for the RL agent to quickly conform to the LLM policy and thereby strengthen the coordination between two agents. The main contributions of this paper can be summarized as follows:

1) **Comprehensive problem formulation.** We present a more comprehensive formulation of the two-stage voltage control problem in ADNs. Compared to existing studies, we explicitly account for the differences in available information between the day-ahead and intra-day stages: in the day-ahead stage, only hourly region-level forecasts are accessible, whereas accurate node-level measurements become available only during the intra-day stage. This setup better reflects the operational conditions of real-world ADNs.

2) **LLM-RL collaboration framework.** To optimize the formulated problem, we propose an LLM-RL collaboration framework that assigns day-ahead OLTC and SCs scheduling to the LLM agent and intra-day PV inverter reactive control to the RL agent. The proposed framework is essentially a hybrid knowledge-data-driven approach that effectively combines the strengths of LLM and RL, not only exploiting the LLM's information processing capabilities and embedded domain knowledge, but also leveraging RL's precise computation and rapid response.

3) **Policy improvement mechanism for LLM and RL.** To enable policy improvement and coordination within the LLM-RL collaboration framework, we also propose dedicated enhancement methods, including a Reflexion-based self-evolution mechanism for the LLM policy and a pretrain-finetune pipeline for the RL policy, enabling efficient adaptation and joint convergence toward optimal performance. Comprehensive comparisons and ablation studies validate the effectiveness and superiority of the proposed method.

II. PRELIMINARIES

*A. Two-Stage Voltage Control Problem Formulation*

In this paper, we consider an ADN whose set of nodes is denoted by $\mathcal{N}$. Based on the spatial boundaries defined by weather forecasts or PV/load predictions, this ADN can be partitioned into $M$ regions. The objective of the two-stage voltage control problem is to minimize voltage deviations across the system and eliminate violations as much as possible:

$$\min \sum_{i \in \mathcal{N}} \sum_{t=0}^{T-1} |V_{i,t} - V_{ref}| \tag{1}$$

where $V_{i,t}$ is the voltage magnitude at node $i$, time $t$; $V_{ref}$ is the voltage reference value; and $T$ is the length of the voltage control process, which is set to one day in this study.

In order to achieve this objective, the ADN operator can dispatch the following controllable equipment: an OLTC installed at the root node; a set of SCs distributed across the network, whose set of nodes is denoted as $\mathcal{N}^{SC}$; and several PV inverters installed at terminal nodes of the network, whose set of nodes is denoted as $\mathcal{N}^{PV}$. We also assume that SCs are not deployed at PV user nodes, i.e., $\mathcal{N}^{SC} \cap \mathcal{N}^{PV} = \emptyset$.

The OLTC serves to raise or lower the overall system voltage through adjustment of its tap position:

$$V_{0,t} = V_{ref} + \beta_t^{OLTC} V_{tap} \tag{2}$$

where $V_{0,t}$ is the root node voltage at time $t$; $V_{tap}$ is the voltage difference between two adjacent tap positions; and $\beta_t^{OLTC}$ is an integer indicating the tap position of the OLTC at time $t$. For example, for an OLTC with 11 tap positions, $\beta_t^{OLTC} \in \{-5, -4, \ldots, -1, 0, 1, \ldots, 4, 5\}$.

The SCs serve to provide reactive power compensation through commitment during periods of peak load, thereby raising the voltage in corresponding local regions:

$$Q_{i,t} = \beta_{i,t}^{SC} Q^{SC} - Q_{i,t}^{LOAD}, \forall i \in \mathcal{N}^{SC} \tag{3}$$

where $Q_{i,t}$ is the reactive injection at node $i$, time $t$; $Q^{SC}$ is the SC capacity; $Q_{i,t}^{LOAD}$ is the reactive load at node $i$, time $t$; and $\beta_{i,t}^{SC} \in \{0,1\}$ is the commitment status of the SC at node $i$, time $t$.

The PV inverters leverage their spare capacity to generate reactive power, thereby refining terminal voltages:

$$Q_{i,t} = Q_{i,t}^{PV} - Q_{i,t}^{LOAD}, \forall i \in \mathcal{N}^{PV} \tag{4}$$

$$|Q_{i,t}^{PV}| \leq \lambda^{PV} \sqrt{(S^{PV})^2 - (P_{i,t}^{PV})^2}, \forall i \in \mathcal{N}^{PV} \tag{5}$$

where $Q_{i,t}^{PV}$ is the reactive generation of the PV inverter at node $i$, time $t$; $S^{PV}$ is the PV installed capacity; $P_{i,t}^{PV}$ is the active generation of the PV inverter at node $i$, time $t$; and $\lambda^{PV} \in [0,1]$ is the reactive capacity factor of the inverter, which is jointly determined by the inverter configurations and ancillary services provided by the PV user.

In addition to the above equipment constraints, the ADN voltage control problem must also satisfy power flow constraints (6)-(9) and voltage limits (10):

$$P_{i,t} = \begin{cases} P_{i,t}^{PV} - P_{i,t}^{LOAD}, \forall i \in \mathcal{N}^{PV} \\ -P_{i,t}^{LOAD}, \forall i \in \mathcal{N} \setminus \mathcal{N}^{PV} \end{cases} \tag{6}$$

$$Q_{i,t} = -Q_{i,t}^{LOAD}, \forall i \in \mathcal{N} \setminus \{\mathcal{N}^{SC} \cup \mathcal{N}^{PV}\} \tag{7}$$

$$P_{i,t} = V_{i,t} \sum_{j \in \mathcal{N}} V_{j,t} (G_{ij} \cos \theta_{ij,t} + B_{ij} \sin \theta_{ij,t}), \forall i \in \mathcal{N} \tag{8}$$

$$Q_{i,t} = V_{i,t} \sum_{j \in \mathcal{N}} V_{j,t} (G_{ij} \sin \theta_{ij,t} - B_{ij} \cos \theta_{ij,t}), \forall i \in \mathcal{N} \tag{9}$$

$$V_{min} \leq V_{i,t} \leq V_{max}, \forall i \in \mathcal{N} \tag{10}$$

where $P_{i,t}$ is the active injection at node $i$, time $t$; $P_{i,t}^{LOAD}$ is the active load at node $i$, time $t$; $G_{ij}$ and $B_{ij}$ are the real and imaginary parts of the admittance element between nodes $i$ and $j$, respectively; $\theta_{ij,t}$ is the voltage phase difference between nodes $i$ and $j$; and $V_{min}$ and $V_{max}$ are the lower limit and upper limit of the voltage magnitude, respectively.

Eqs. (1)-(10) represent the mathematical modeling of the two-stage voltage control problem adopted by most existing studies. However, in real-world ADN operations, particularly in scheduling mechanical assets, a series of grid codes must be followed, as the control of these devices often involves temporal coupling constraints and requirements for manual on-site intervention. In this paper, we consider two common grid code requirements for OLTC and SCs, i.e., "The number of daily OLTC adjustments is capped by an upper limit," and "SCs can only be committed during a specific time window each day." As can be seen, these grid code requirements can be easily expressed in natural language, yet effectively embedding them into a data-driven RL policy requires complicated design. This is one of the main motivations for introducing LLMs, as LLMs can readily comprehend such operational constraints and generate compliant scheduling strategies.

The solution to the above constrained optimization problem is usually carried out in two stages, i.e., the day-ahead stage and the intra-day stage. In the day-ahead stage, the ADN operator obtains hourly region-level forecasts and determines scheduling strategies for OLTC and SCs to proactively regulate the overall voltage profile for the following day, preventing severe voltage violations. Then in the intra-day stage, real-time accurate node-level measurements become accessible, and the ADN operator leverages the fast response capability of PV inverters to further refine terminal voltages.

*B. LLM as Decision-Making Models*

In recent years, the rapid development of LLMs has become one of the most transformative and significant developments in artificial intelligence, bringing profound impacts to numerous specialized domains. Built upon the transformer architecture [26] and trained on vast amounts of pretraining data, existing LLMs have acquired powerful information processing capabilities and embedded domain knowledge, making it possible to employ them as decision-making models. Some advanced research in artificial intelligence has begun to apply LLMs as decision-making models in autonomous driving [27], [28] and robotic control [29], [30]. These models take text-based or multimodal environmental descriptions as input and generate corresponding control strategies in an end-to-end manner, offering a novel pathway for solving complex decision-making problems.

In power systems, dispatch problems, such as ADN voltage control, are quintessential decision-making problems. However, existing applications of LLMs in the power domain have largely focused on tasks like simulation setup [31], [32], document analysis [33], and scenario generation [34]. The few studies related to dispatching typically assign LLMs subtasks such as code programming and function synthesis, which remain fundamentally natural-language processing problems and do not fully exploit the capabilities of existing LLMs. Therefore, in this paper, we innovatively introduce the LLMs as decision-making models and apply them to the ADN voltage control problem. Through the proposed input and output formulation, LLM-RL collaboration framework, and policy improvement mechanisms, the LLM agent successfully addresses the challenge of coordinating mechanical assets under grid code constraints and enhances voltage control performance in ADNs.

*C. MDP and RL*

The intra-day strategies are provided by the RL agent. In RL, the sequential decision-making process is usually modeled as a Markov decision process (MDP) [35], which can be described by a tuple $(\mathcal{S}, \mathcal{A}, p, r, \gamma)$, where $\mathcal{S}$ is the state space of the environment; $\mathcal{A}$ is the action space of the RL agent; $p$ is the state transition function; $r$ is the reward function; and $\gamma \in [0,1)$ is the discount factor for future rewards.

At time $t$, the RL agent receives the state observation $s_t \in \mathcal{S}$ of the current environment and decides its control action $a_t \in \mathcal{A}$ based on its policy $\pi: \mathcal{S} \times \mathcal{A} \to [0, \infty)$, i.e., $a_t \sim \pi(\cdot | s_t)$. Then the environment transfers to the next state $s_{t+1} \in \mathcal{S}$ based on $p: \mathcal{S} \times \mathcal{A} \times \mathcal{S} \to [0, \infty)$ and gives the RL agent a reward $r_t(s_t, a_t)$ as feedback. Using the feedback information from the environment, the objective of the RL agent is to optimize its policy $\pi$ so that the expected cumulative rewards $J(\pi)$ can be maximized, i.e.,

$$\max_\pi J(\pi) = \mathbb{E}_{p,\pi}\left[\sum_{t=0}^{T-1} \gamma^t r_t\right] \quad (11)$$

Detailed MDP settings of the voltage control problem and RL training mechanism will be described in the next section.

## III. METHODS

*A. LLM-RL Collaboration Framework*

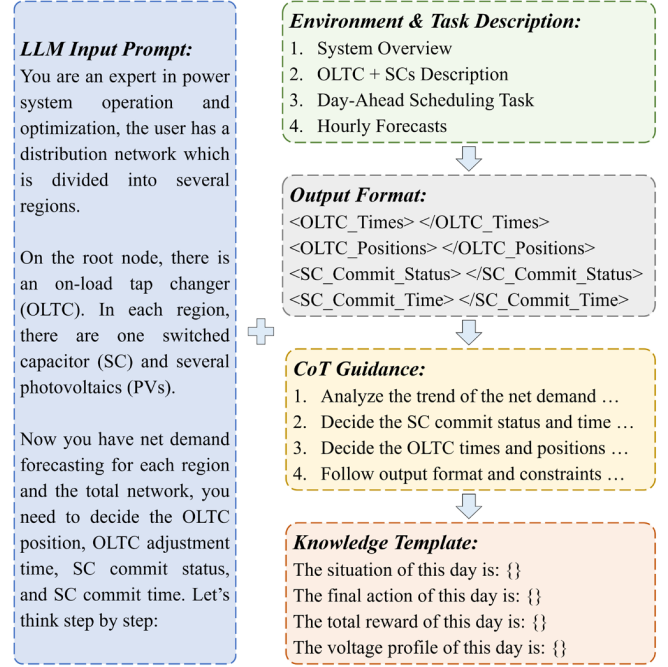

**Fig. 2.** Illustrative prompts for LLM strategies generation.

As shown in Fig. 1, the proposed LLM-RL collaboration framework employs an LLM agent and an RL agent to handle day-ahead and intra-day decision-making, respectively, achieving voltage optimization through their collaborative interaction. In the day-ahead stage, the input to the LLM agent consists of hourly region-level forecasts, specifically the net load forecasts for the $M$ regions and the overall ADN system. These inputs are characterized by coarse granularity and may exhibit certain deviations due to the limited accuracy of current PV/load predictions. Based on this information, the LLM agent outputs scheduling strategies for the OLTC and SCs that comply with grid code requirements. Then in the intra-day stage, the input to the RL agent consists of accurate node-level measurements, specifically the real-time loads and PV generation. These inputs are characterized by high resolution and relatively high accuracy. Based on this information, the RL agent outputs reactive control strategies for the PV inverters that comply with reactive capacity constraints. The key innovation of this collaboration framework lies in the introduction of the LLM agent, which enables the exploitation of its advantages in information processing and knowledge integration.

To adapt an existing general-purpose LLM into the required domain-specific LLM agent, that is, one capable of accurately understanding the voltage control problem and generating day-ahead OLTC and SCs scheduling strategies that satisfy operational constraints, we propose the tailored prompts. As illustrated in Fig. 2, the prompts consist of the following components:

1) *Environment and task description:* This component provides a detailed description of the environment and task, including the system overview, descriptions of the OLTC and SCs, and a clear specification of the day-ahead scheduling task, which requires the LLM agent to generate day-ahead strategies based on the hourly forecasts for the following day.

2) *Output format:* A standardized output format can effectively reduce the occurrence of hallucinations or internal errors. In this component, we specify the required output format for the LLM agent, namely, the timing and corresponding tap positions for OLTC adjustments, as well as the commitment statuses and time windows for SCs. Once the LLM agent generates the content, we apply regular expressions to extract structured scheduling strategies from the response and validate their compliance with grid code requirements. If the strategies satisfy these operational constraints, the policy is deemed valid; otherwise, the LLM agent is prompted to regenerate the day-ahead scheduling strategies in another dialogue.

3) *Chain-of-Thought (CoT) guidance:* CoT reasoning refers to decomposing a complex problem into a sequence of simple, executable inference steps [36]. For domain-specific problems, CoT can effectively assist in generating reliable outputs. Specifically, in the context of voltage control, the LLM agent is first instructed to analyze the trend and magnitude of net load forecasts; based on which it determines the commitment actions for SCs in each region and the tap positions for the OLTC; and finally, it organizes these decisions into a formal final answer which returns to the ADN operator.

4) *Few-shot examples as knowledge embedding:* The above approach only ensures that the generated strategies are feasible, but it does not enable improvement of the LLM policy. To realize policy improvement, we further construct a knowledge base $\mathcal{D}_{LLM}$ that stores information from several historical days. For a specific historical day, we store its day-ahead forecasting data, generated OLTC and SCs actions, and the resulting total reward and voltage profiles. When making decisions, the LLM agent retrieves the historical day most similar to the current scenario from the knowledge base and embeds it as a few-shot example in the prompts for reference during reasoning. This prevents the LLM agent's decision-making from being arbitrary and enables continuous policy optimization through iterative updates of the knowledge base. Complete prompts for the LLM agent are provided in an online supplementary file [37]. And the details of LLM policy improvement, historical knowledge retrieval, and knowledge base update will be elaborated in the following subsections.

On the other hand, the role of the RL agent is to adapt the OLTC and SCs settings prescribed by the LLM agent in the day-ahead stage and further optimize the terminal voltages during the intra-day stage by adjusting the reactive power generation of PV inverters. Although the RL policy can be readily improved using off-the-shelf RL algorithms, rapidly adapting it to the LLM policy within the proposed LLM-RL collaboration framework poses a significant challenge. Because when the statuses of the OLTC and SCs change, the RL agent actually faces a different environment. Therefore, the training mechanism of the RL agent should be carefully designed to enhance its learning capability and adaptation rate.

As shown in Fig. 3, to address this issue, we further propose a pretrain-finetune pipeline for the RL agent. Specifically, the RL agent first undergoes extensive exploration in a pretraining environment to learn a general control policy for PV inverters. Subsequently, the pretrained RL agent is transferred to a finetuning environment, where it adapts to the current LLM policy. Notably, in the pretraining environment, the actions of OLTC and SCs are randomly assigned, whereas in the finetuning environment, these actions are provided by the well-improved LLM agent. This design enables the RL agent to acquire versatile control strategies across diverse settings during pretraining, thereby facilitating rapid adaptation to the LLM policy during finetuning.

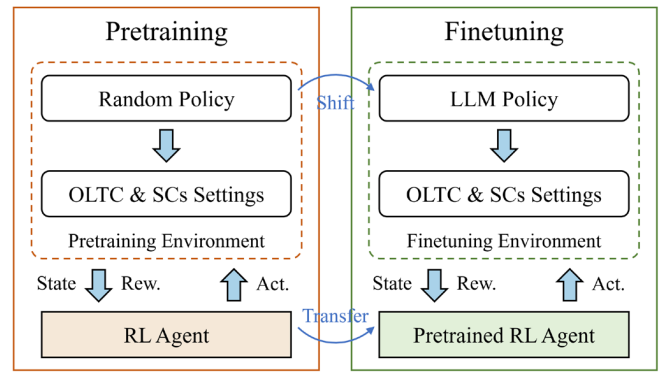

**Fig. 3.** Pretrain-finetune pipeline for RL agent.

*B. Intra-Day PV Reactive Control Policy Improvement*

Within the pretrain-finetune pipeline, in order to leverage RL algorithms to improve the intra-day PV reactive control policy, the intra-day voltage control problem should first be formulated as an MDP. Corresponding definitions are designed as follows:

1) *State space:* The state $s_t \in \mathcal{S}$ of the MDP consists of two

components, i.e., the real-time node-level measurements ($P_t^{LOAD}, Q_t^{LOAD}, V_t, P_t^{PV}, Q_t^{PV}$) from the ADN monitoring system and OLTC and SCs actions ($\beta_t^{OLTC}, \beta_t^{SC}$) previously decided by the LLM agent, where $P_t^{LOAD}, Q_t^{LOAD}, V_t$ are the vectors of the ADN's active loads, reactive loads, and voltage magnitudes; $P_t^{PV}, Q_t^{PV}$ are the vectors of the PVs' active and reactive generation; and $\beta_t^{SC}$ is the vector of the SCs' commitment statuses. Specifically, OLTC and SCs actions ($\beta_t^{OLTC}, \beta_t^{SC}$) are embedded into the state space using one-hot encoding. Incorporating the LLM agent's actions into the state space enhances the RL agent's environmental awareness and improves its adaptation to the LLM policy.

2) *Action space:* The action $a_t \in \mathcal{A}$ of the MDP is constructed with the reactive generation $Q_t^{PV}$ of all controllable PV inverters in the ADN. Specifically, we constrain the action of the RL agent in $[-\lambda^{PV}, \lambda^{PV}]$ so that the reactive capacity constraint eq. (5) is always satisfied.

3) *Reward function:* The reward function of the MDP aligns with the objective of the voltage control problem:

$$r_t = -\frac{\sum_{i \in \mathcal{N}} |V_{i,t} - V_{ref}|}{|\mathcal{N}|} \quad (12)$$

The RL algorithm utilized in this paper is proximal policy optimization (PPO) proposed in [38]. In PPO, the RL agent's policy $\pi$ is approximated by a neural network $\pi_\theta$ with parameters $\theta$. After completing one episode of iteration between the old policy $\pi_{\theta'}$ and the external environment, the objective function for $\pi_\theta$ can be expressed as:

$$J_{\theta'}(\theta) = \mathbb{E}_{(s_t,a_t)\sim\pi_{\theta'}}[\rho_t(\theta) A_{\theta',t}] \quad (13)$$

where $(s_t, a_t) \sim \pi_{\theta'}$ is the batch of iteration data generated from the old policy $\pi_{\theta'}$; $\rho_t(\theta) = \pi_\theta(a_t|s_t)/\pi_{\theta'}(a_t|s_t)$ is the importance sampling ratio; and $A_{\theta',t}$ is an estimator of the advantage function using generalized advantage estimator (GAE) technique.

In order to constrain the updated policy within a trust region and ensure more stable training, a common practice in PPO is to use gradient clipping, i.e.,
$J_{\theta'}^{CLIP}(\theta) = \mathbb{E}_{(s_t,a_t)\sim\pi_{\theta'}}$
$[\min(\rho_t(\theta) A_{\theta',t}, clip(\rho_t(\theta), 1 - \varepsilon_{PPO}, 1 + \varepsilon_{PPO}) A_{\theta',t})] \quad (14)$
where the clip function confines $\rho_t(\theta)$ to the interval $[1 - \varepsilon_{PPO}, 1 + \varepsilon_{PPO}]$ so that the updated policy does not deviate excessively from the old policy. Then based on the calculated $J_{\theta'}^{CLIP}(\theta)$, the policy network $\pi_\theta$ is optimized $n_{PPO}$ times by updating its parameters.

*C. Day-Ahead OLTC and SCs Scheduling Policy Improvement*

After ensuring that the LLM agent can comprehend the problem and generate day-ahead strategies that comply with operational constraints, the next consideration is how to improve the LLM policy accordingly. In this paper, we propose a Reflexion-based self-evolution mechanism for the LLM agent. Similar to RL algorithms, Reflexion enhances the agent's policy through interaction with an environment and feedback from it. Specifically, Reflexion aims to improve the model's reasoning and decision-making capabilities on complex tasks via a self-reflection process that encompasses execution, evaluation, reflection, memory, and iterative refinement.

For the voltage control problem discussed in this paper, a knowledge base $\mathcal{D}_{LLM}$ is firstly maintained for the LLM agent, storing data from several historical days. When a decision is required for a new day $d_i$, the LLM agent queries the knowledge base and retrieves the historical scenario most similar to current situation (with the highest similarity $\hat{\varepsilon}$), which is then embedded as a few-shot example into the prompts provided to the LLM agent. Specifically, for the new day $d_i$ and a historical day $d_j$, their similarity $\varepsilon_{ij}$ is calculated as follows:

$$\varepsilon_{ij} = TS_{ij} \times MS_{ij} \quad (15)$$

where $TS_{ij}$ captures the temporal similarity between $d_i$ and $d_j$; and $MS_{ij}$ captures the magnitude similarity between $d_i$ and $d_j$. $TS_{ij}$ and $MS_{ij}$ are designed as follows:

$$TS_{ij} = \frac{\sum_{t=0}^{23} \widetilde{LOAD}_{i,t} \times \widetilde{LOAD}_{j,t}}{\sqrt{\sum_{t=0}^{23}(\widetilde{LOAD}_{i,t})^2} \times \sqrt{\sum_{t=0}^{23}(\widetilde{LOAD}_{j,t})^2}} \quad (16)$$

$$MS_{ij} = \frac{\min(|\sum_{t=0}^{23} \widetilde{LOAD}_{i,t}|, |\sum_{t=0}^{23} \widetilde{LOAD}_{j,t}|)}{\max(|\sum_{t=0}^{23} \widetilde{LOAD}_{i,t}|, |\sum_{t=0}^{23} \widetilde{LOAD}_{j,t}|)} \quad (17)$$

where $\widetilde{LOAD}_{i,t}$ ($t \in [0,23]$) is the hourly net load forecast for the overall ADN system in $d_i$.

After the OLTC and SCs actions are executed in the training environment, the LLM agent then receives the corresponding results and feedback, including the total reward of the day and hourly voltage profiles, i.e., average voltage for each region and the overall ADN system. Based on the feedback and following the Reflexion paradigm, we instruct the LLM agent to refine its original actions. For example, if the overall voltage is too low during a specific time period, the LLM agent will increase the OLTC tap position for the corresponding period. After performing $n_{LLM}$ rounds of Reflexion, we obtain $n_{LLM} + 1$ candidate rewards for the new day $d_i$, and we denote the highest one as $R_i$.

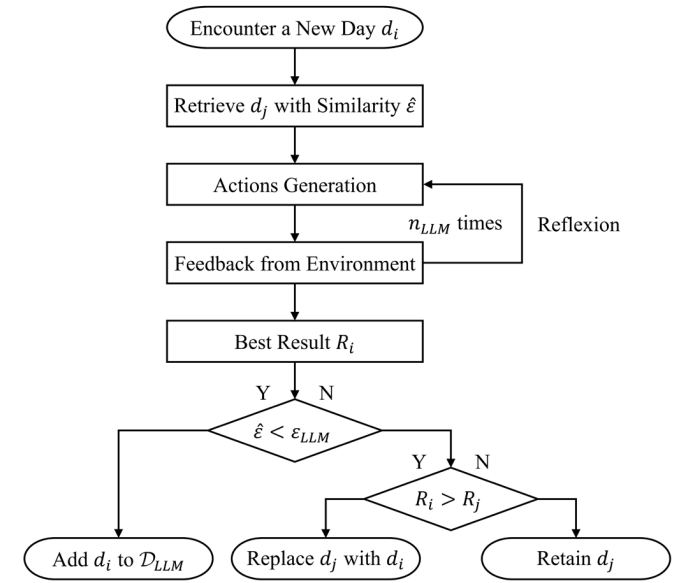

**Fig. 4.** Knowledge base updating process for LLM agent.

After Reflexion concludes, the next step is to update the knowledge base $\mathcal{D}_{LLM}$, where we set an updating threshold

$\varepsilon_{LLM}$. Then the update process should be handled in two cases: If $\hat{\varepsilon} < \varepsilon_{LLM}$, it indicates that the new day $d_i$ is a novel situation which have not been encountered before, then the new day $d_i$, the reward $R_i$, and its corresponding OLTC and SCs actions are directly appended into $\mathcal{D}_{LLM}$. If $\hat{\varepsilon} \geq \varepsilon_{LLM}$, it indicates that $d_i$ has been encountered before and can be partially represented by a historical day $d_j$. We then compare the total rewards $R_i$ and $R_j$ to examine their performance. If $R_i > R_j$, it means the LLM agent has discovered a better set of actions for this situation, and we replace $d_j$ in $\mathcal{D}_{LLM}$ with $d_i$. Conversely, if $R_i \leq R_j$, it implies that the previously stored day $d_j$ corresponds to a superior strategy, thus $d_j$ is retained in $\mathcal{D}_{LLM}$ and $d_i$ is discarded. This updating process is also illustrated in Fig. 4.

Drawing an intuitive analogy between the proposed Reflexion-based self-evolution method and RL algorithms, the knowledge base $\mathcal{D}_{LLM}$ and the retrieved historical day $d_j$ serves as the policy parameters $\theta$, instructing the agent to generate corresponding actions; the Reflexion process is analogous to the optimization process of $J_{\theta'}^{CLIP}(\theta)$, where the agent identifies directions for policy improvements; $\varepsilon_{LLM}$ is like $\varepsilon_{PPO}$, which constrains differences between the old policy and the updated one; and $n_{LLM}$ is like $n_{PPO}$, which denotes the number of times an episode of data can be reused. Through the iterative Reflexion process and updates of the knowledge base, the LLM policy can be progressively refined. In addition, it should be noted that Reflexion is essentially a "training" process based on iterative experiences, and thus it occurs only in an offline manner. During online execution, the knowledge base is frozen and the LLM directly retrieves the most similar historical scenario from the knowledge base for inference, as it is infeasible and impractical for online Reflexion.

## IV. NUMERICAL STUDIES

### A. Cases and Methods Setup

In this section, numerical simulations are conducted on IEEE 33-bus [39] and 141-bus [40] distribution systems. OLTC at root node has a total of 11 tap positions, each corresponding to a voltage regulation step of 0.6%. Voltage limitations are set at [0.95, 1.05] p.u. The 33-bus system is partitioned into three regions, and the 141-bus system is partitioned into five regions, in which each region is equipped with one SC and two PV inverters. We use one year of PV and load data to simulate power flow in the distribution systems. In the day-ahead stage, the LLM agent receives hourly region-level forecasts, to which we add 5% white noise to emulate the prediction errors in current PV and load predictions. For the RL agent, intra-day control interval is 15 minutes and the length of the control process is 96 steps. Corresponding configurations are listed in Table I, and more detailed specifications and topologies are provided in the online supplementary file [37].

As for the LLM and RL used in this paper, considering the randomness in LLM responses and the PPO algorithm, we use 3 independent random seeds for each experiment. All experiments are run on a computer with a 2.2GHz Intel Core i9-14900HX CPU and a 32GB RAM. Hyperparameters for the LLM agent and the RL agent are presented in Table II.

TABLE I
CONFIGURATIONS OF THE TEST SYSTEMS

| Parameter | Value |
|---|---|
| No. Region | 3, 5 |
| No. SC | 3, 5 |
| No. PV | 6, 10 |
| No. OLTC Position | 11 |
| OLTC Adjustment Limits | 4 |
| OLTC Step | 0.6% |
| SC Capacity | 0.15MVAR, 0.4MVAR |
| $\lambda^{PV}$ | 0.3, 0.15 |

For comparison and ablation studies, in order to demonstrate the effectiveness of the proposed approach, in the offline training phase, we set the proposed method against three baselines: 1) Pure-RL, in which the day-ahead and intra-day policies are optimized by two separate RL agents, is designed to validate the effectiveness of the LLM-RL collaboration framework. To ensure compliance with grid code requirements, we further incorporate a soft constraint and a hard constraint into the day-ahead scheduling strategies of Pure-RL, following established practices in the literature. The soft constraint refers to the penalty mechanism: when the RL agent proposes an action that violates grid code limits, it receives a corresponding negative penalty; while the hard constraint refers to action masking, which enforces that the states of the OLTC and SCs remain unchanged once their maximum allowable switching counts are reached. It should be noted that although Pure-RL receives a penalty term as feedback during training, this penalty term is excluded from the reward metrics presented in the following results to ensure a fair comparison. 2) No-PT, in which the RL agent is not pretrained and is directly finetuned in the environment defined by the LLM policy, is designed to validate the necessity of the proposed pretrain-finetune pipeline. 3) No-Reflexion: in which the Reflexion process is omitted, is designed to demonstrate the effectiveness of the proposed Reflexion-based mechanism in complex tasks.

TABLE II
HYPERPARAMETERS OF THE LLM AGENT AND THE RL AGENT

| Agent | Parameter | Value |
|---|---|---|
| LLM | Model | Qwen-Plus |
| | Version | "2025-12-01" |
| | Temperature (Training) | 0.7 |
| | Top-P (Training) | 0.8 |
| | Temperature (Test) | 0.2 |
| | Top-P (Test) | 0.6 |
| | $\varepsilon_{LLM}$ | 0.7 |
| | $n_{LLM}$ | 3 |
| RL | Mini Batch Size | 32 |
| | Hidden Layers | 2 |
| | Hidden Units | 256 |
| | $\gamma$ | 0.9 |
| | Learning Rate (Pretrain) | 1e-4 → 1e-5 (Linear Decay) |
| | Learning Rate (Finetune) | 1e-5 |
| | $\varepsilon_{PPO}$ | 0.3 |
| | $n_{PPO}$ | 4 |

Then, in the online execution phase, in addition to the aforementioned methods, we also design another three baselines: 1) Original, in which the OLTC is fixed at the reference voltage position, SC commitment is disabled, and PV inverters are prohibited from participating reactive control, and this baseline is used to demonstrate the necessity of proactive voltage control measures in ADNs. 2) No-LLM, in which the mechanical assets OLTC and SCs are disabled, is designed to validate their capability of regulating the overall voltage profile. 3) No-RL, in which the PV reactive generation is disabled, is designed to validate its capability of refining terminal voltages.

*B. Training Performance of Proposed and Baseline Methods*

To compare the performance of Proposed, Pure-RL, No-PT, and No-Reflexion during the offline training phase, we plot their training curves in Figs. 5-6. Solid lines in the figures represent the mean values across multiple random seeds, while the shaded areas indicate the corresponding error bounds. Also, considering that voltage control performance may vary significantly over the course of a year, we apply a moving average with a window length of 25 episodes to smooth these curves.

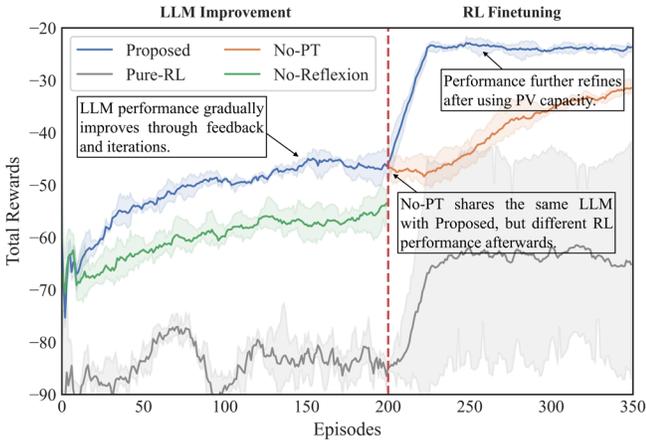

**Fig. 5.** Training performance in IEEE 33-bus system.

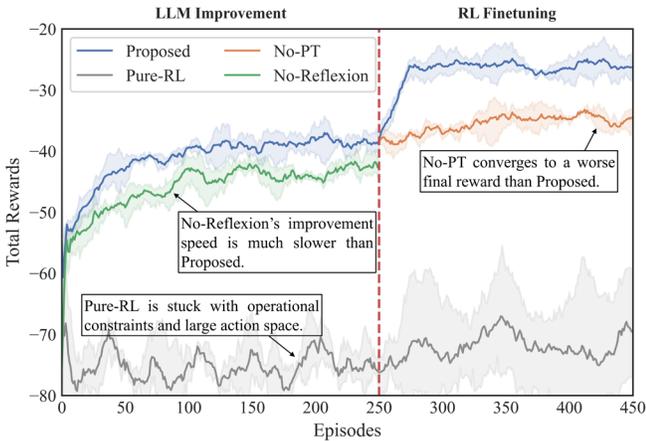

**Fig. 6.** Training performance in IEEE 141-bus system.

As can be seen, the training phase begins with the LLM improvement; once the LLM policy stabilizes, we freeze its knowledge base and introduce a pretrained RL agent into the environment for RL finetuning. In the 33-bus system, the LLM agent undergoes improvement over 200 episodes, while the RL agent is pretrained for 1500 episodes and finetuned for 150 episodes. In the 141-bus system, the LLM agent's improvement spans 250 episodes, and the RL agent is pretrained for 2500 episodes and finetuned for 200 episodes. In addition, for better visualization and comparison, we present the complete training curves of Proposed and Pure-RL in the figures, but only show the first half of the training curve for No-Reflexion and the second half for No-PT. However, it should be noted that all these four methods undergo the complete LLM improvement and RL finetuning procedures, and their converged performance will be compared in the next subsection.

First, examining the training curve of Proposed reveals that, during the first half, the LLM policy progressively improves and converges steadily through interaction with the environment and continuous updates of the knowledge base, demonstrating the effectiveness of the proposed Reflexion-based self-evolution mechanism. In the second half of the curve, after the RL agent is introduced, the reward further improves using the PV reactive generation capacity. Moreover, since the RL agent has been well-pretrained beforehand, it can quickly adapt to the day-ahead strategies provided by the LLM agent, leading to stable convergence and enabling efficient coordination between day-ahead and intra-day operations.

Second, we compare the performance of Pure-RL. It can be observed that, even without iteration and improvement, the initial LLM policy already outperforms the RL policy on the day-ahead voltage scheduling task. This is because the LLM possesses extensive embedded knowledge and, guided by the proposed prompts, can generate compliant and reasonable strategies. For example, it lowers the OLTC tap position when PV generation is high and dispatches SCs during peak load periods. Such intuitive operational knowledge is largely absent in the RL agent, resulting in an initial policy that is relatively random and ineffective. And even after interacting with the environment, Pure-RL shows no obvious performance improvement. On the one hand, the RL agent must handle operational constraints and a large action space, requiring substantially more exploration than the LLM agent to develop an effective understanding of the environment. On the other hand, the feedback received by the RL agent is a simple scalar reward, whereas the LLM agent can process heterogeneous information, such as voltage profiles and system states, which provides much richer contextual cues. Compared to a scalar reward, this diverse information offers the LLM agent clear directions for policy improvement, while the RL policy remains largely stochastic.

In summary, on the day-ahead voltage scheduling task, the LLM demonstrates at least three advantages over RL, i.e., embedded knowledge driven strategy generation, semantic comprehension enhanced constraint handling, and reasoning backed policy improvement. These advantages underscore both the superiority and necessity of applying LLMs to specialized dispatch tasks, as well as the effectiveness of the proposed LLM-RL collaboration framework.

Third, we observe the training curve of No-PT. Since the only difference between No-PT and Proposed is whether the introduced RL agent has undergone pretraining, and both methods share the same LLM agent, their curves are identical in the first half and diverge only in the second half. It can be found that, due to the absence of pretraining, the RL agent in No-PT learns and converges more slowly than that in Proposed in the 33-bus system. And in the 141-bus system, it even converges to a suboptimal policy. This demonstrates the effectiveness of the proposed pretrain-finetune pipeline. The pretraining process not only provides the RL agent with a high-quality warm start policy but also enhances its ability to adapt to diverse OLTC and SCs settings. Therefore, pretraining the RL agent is essential within the proposed LLM-RL collaboration framework.

Finally, an analysis of No-Reflexion reveals that, although it is also capable of self-evolution through interaction with the environment, its improvement rate is significantly lower than that of Proposed. This is because No-Reflexion does not perform in-depth reasoning and reflection on environmental feedback and instead relies on trial-and-error learning. We still use RL algorithms by analogy, which in fact indicates that the interaction samples of No-Reflexion are not fully utilized, thereby reducing "sample efficiency" and degrading training performance.

*C. Execution Performance of Proposed and Baseline Methods*

After the LLM agent and the RL agent are fully trained and converged, the system enters the online execution phase. During this phase, the LLM agent's knowledge base and the RL agent's policy parameters are no longer updated, and Reflexion is disabled. Instead, day-ahead and intra-day strategies are generated directly. In real-world ADNs, the latest knowledge base and policy parameters are stored and maintained by the ADN operator. In the day-ahead stage, the LLM agent determines the scheduling of OLTC and SCs based on forecast information, whose strategies are generated within minutes. Then during the intra-day operation, the RL agent determines the reactive power generation based on real-time measurements, with strategy generation occurring in milliseconds, which is totally sufficient to meet the requirements of real-time control.

To evaluate the execution performance of the proposed and baseline methods, we select 25 additional episodes for each random seed to test their converged policies, resulting in a total of 75 test episodes per method. We present the tested average nodal voltage deviations and the nodal voltage violation rates in Tables III and IV, in which the best performance is marked in bold.

TABLE III
EXECUTION PERFORMANCE OF IEEE 33-BUS SYSTEM

| Method | Voltage Deviation (p.u.) | | Violation Rate (%) | |
|---|---|---|---|---|
| | Mean | Std. | Mean | Std. |
| Original | 2.50e-02 | 4.95e-03 | 10.7 | 6.06 |
| No-LLM | 1.14e-02 | 2.48e-03 | 3.65e-02 | 1.62e-01 |
| No-RL | 1.69e-02 | 3.18e-03 | 1.21 | 2.79 |
| Proposed | **8.22e-03** | 2.08e-03 | **2.60e-03** | **1.38e-02** |
| Pure-RL | 2.06e-02 | 7.09e-03 | 3.24 | 5.31 |
| No-PT | 1.02e-02 | 2.28e-03 | 1.26 | 1.53 |
| No-Reflexion | 8.60e-03 | **1.97e-03** | 6.08e-03 | 2.42e-02 |

* Mean and Std. denote mean values and standard deviations across 75 episodes.

TABLE IV
EXECUTION PERFORMANCE OF IEEE 141-BUS SYSTEM

| Method | Voltage Deviation (p.u.) | | Violation Rate (%) | |
|---|---|---|---|---|
| | Mean | Std. | Mean | Std. |
| Original | 2.16e-02 | 3.35e-03 | 3.74 | 4.82 |
| No-LLM | 1.21e-02 | 3.23e-03 | 1.46 | 2.87 |
| No-RL | 1.21e-02 | 3.26e-03 | 1.35e-01 | 3.64e-01 |
| Proposed | **8.37e-03** | **2.21e-03** | **1.58e-02** | **8.48e-02** |
| Pure-RL | 2.30e-02 | 6.30e-03 | 2.89 | 3.46 |
| No-PT | 1.11e-02 | 2.89e-03 | 7.67e-01 | 1.59 |
| No-Reflexion | 1.00e-02 | 2.33e-03 | 4.70e-02 | 3.20e-01 |

* Mean and Std. denote mean values and standard deviations across 75 episodes.

As shown in Tables III and IV, Original exhibits the worst performance on the voltage control problem, resulting in large voltage deviations and frequent voltage violations, which demonstrate the necessity of proactive control in ADNs. Although No-LLM and No-RL show significant improvements over Original, their performance still falls short of that of Proposed. This not only confirms the effectiveness of the proposed method but also underscores the importance of coordinating mechanical assets with flexible resources. Especially in the context of widespread integration of distributed energy resources into ADNs, coordinated control of multiple equipment has become a critical approach for addressing voltage challenges.

The test results of Pure-RL show that, even at the final stage of training, it still fails to discover a high-quality control policy and performs worse than Original in certain scenarios. This again validates the effectiveness of incorporating the LLM agent, which can significantly improve both optimization efficiency and solution quality on specific problems. As for No-PT and No-Reflexion, their performance is consistent with their training behavior, i.e., the absence of pretraining and Reflexion process leads to performance degradation compared to Proposed. These comparisons also demonstrate the effectiveness of the proposed pretrain-finetune pipeline and the Reflexion-based self-evolution mechanism.

## V. CONCLUSION

To address voltage challenges in ADNs caused by high PV penetration and the inability of existing data-driven methods to effectively handle heterogeneous information, semantic constraints, and dynamic instructions, we propose a hybrid LLM-RL collaboration framework for two-stage voltage control: the LLM agent schedules OLTCs and SCs using day-ahead forecasts and historical records, while the RL agent adjusts PV inverter reactive power based on real-time measurements. Building upon this framework, we further introduce a Reflexion-based self-evolution mechanism tailored for the LLM agent and a pretrain-finetune pipeline designed for the RL agent, enabling effective policy enhancement and efficient coordination between these two agents. Experimental results show that our approach significantly outperforms baseline methods in both voltage regulation performance and training efficiency, demonstrating the effectiveness and necessity of synergizing knowledge-driven reasoning with data-driven control.

Drawing from our research and operational experience, we found that LLMs excel at knowledge-based reasoning and generation but are less capable in precise numerical computation. This is also one of the key motivations for proposing the LLM-RL collaboration framework. In future work, how to fully leverage the strengths of LLMs and achieve deeper integration with various domain-specific models represents an important and promising research direction.